# Low-cost IoT-Based Rainfall Monitoring with Web-Based Data Access

Mohammad Solaiman[a], Ronan Reza[b], Su Zhang[b,c,d], Gerhard Schoener[b], Fernando Moreu[b*]

[a]*Department of Electrical and Computer Engineering, University of New Mexico, 1 University of New Mexico, Albuquerque, NM 87131, USA*

[b]*Department of Civil, Construction and Environmental Engineering, University of New Mexico, 1 University of New Mexico, Albuquerque, NM 8713, USA*

[c]*Department of Geography and Environmental Studies, University of New Mexico, 1 University of New Mexico, Albuquerque, NM 8713, USA*

[d]*Earth Data Analysis Center, University of New Mexico, 400 University Blvd NE Albuquerque, NM 87106, USA*

** Corresponding author. Email: fmoreu@unm.edu*

## Abstract

Rainfall measurement with high spatial and temporal resolution is critical for flood forecasting, drought mitigation, and disaster preparedness. Rainfall patterns are highly variable, both geographically and over time. This variability presents a significant challenge for monitoring, as rain gauges can accurately capture temporal patterns only at a single location. Furthermore, the high cost of commercial instruments restricts their widespread deployment, and rain gauge networks often fail to adequately capture the spatial heterogeneity of precipitation patterns. To address these limitations, this study introduces a low-cost IoT-based rainfall monitoring system developed upon the Low-cost Efficient Wireless Intelligent Sensor (LEWIS) platform. Four rainfall sensors were designed, developed, and deployed at different locations across the semi-arid region of the United States, in the State of New Mexico, to capture localized precipitation variability. Each sensor node integrates a rainfall detection module with an LTE-enabled microcontroller and is powered by a compact solar-battery system, ensuring autonomous and self-sufficient operation. Real-time precipitation data are transmitted to a cloud server for continuous access, visualization, and integration with early-warning frameworks. The results demonstrate that IoT-based rainfall monitoring can achieve reliable accuracy at a fraction of the cost of conventional gauges, while supporting dense deployment for microscale precipitation analysis. Comparative validation with model-based precipitation data and in situ observations shows strong agreement in the detection and timing of recorded precipitation events, highlighting the system's potential for early warning, disaster risk reduction, and bias correction of remotely sensed precipitation products by filling observational gaps in under-instrumented semi-arid areas.

---

## 1. Introduction

Rainfall is a key driver of the hydrological cycle, regulating runoff, infiltration, groundwater recharge, and vegetation growth (Sala and Lauenroth, 1982). It plays a critical role in maintaining ecological balance, agricultural productivity, and water resource availability (Ghebreyesus and Sharif, 2020; Qiu et al., 2025). Variations in rainfall intensity and frequency influence flood risks, drought severity, and long-term water supply (Ding et al., 2020; Thomas et al., 2016). In arid and semi-arid regions such as the State of New Mexico, rainfall is highly variable, typically manifesting as short, intense events alternating with prolonged dry periods (Chen et al., 2022; Vargas Godoy et al., 2025). This irregularity presents considerable challenges for water management, soil conservation, and agricultural planning. Therefore, continuous, high-resolution monitoring of rainfall in both space and time is essential for informing effective adaptation and mitigation strategies (Paradkar and Mittal, 2024; Santana et al., 2023).

Effective rainfall monitoring is widely recognized as essential for improving hydrological modeling and water resource management, given the strong spatial and temporal variability of precipitation. Studies have shown that coarse or sparsely distributed rainfall observations can introduce substantial errors in runoff prediction, soil moisture estimation, and flood assessment, particularly in dryland environments where storms are highly localized (Rezaei et al., 2011; Sorensen et al., 2024). Research in semi-arid regions, including the U.S. Southwest, indicates that conventional monitoring networks often fail to accurately capture short-duration convective storms, resulting in underestimation of peak rainfall intensity and accumulated precipitation totals. Consequently, understanding rainfall distribution at finer spatial and temporal scales is critical for reliable hydrological modeling, water resource planning, and climate forecasting.

Rainfall variability is also a key indicator of climate change, as global warming intensifies the hydrological cycle and alters precipitation patterns (Fowler et al., 2021; Wang and Liu, 2023). Many regions are now experiencing extreme rainfall events, while others face prolonged droughts (Gründemann et al., 2025; Trenberth, 2005). These fluctuations have direct implications for food security, urban drainage systems, and natural ecosystems (Njouenwet et al., 2022). Reliable rainfall data are crucial not only for short-term applications, such as flood forecasting and irrigation management, but also for long-term trend analysis and climate modeling (Khayyat, 2022; Pandey, 2023). Ground-based observations remain essential for calibration and validation of these models; however, dense networks of traditional gauges are often constrained by high costs and logistical challenges (Schmeller et al., 2022). Conventional rainfall measurement systems, such as tipping bucket gauges, weighing gauges, and optical sensors, have been widely used for decades due to their accuracy and simplicity (Lanza et al., 2022). Nevertheless, these instruments are expensive to install and maintain, and their mechanical components are prone to wear, clogging, or drifting under field conditions (Segovia-Cardozo et al., 2023). Weighing gauges additionally require frequent calibration and stable power sources. Consequently, many rural and remote regions remain under-instrumented, creating substantial data gaps that hinder hydrological modeling and early warning capabilities. The high operational cost, limited scalability, and maintenance demands of conventional systems highlight the necessity of developing new, affordable, and reliable alternatives capable of long-term autonomous operation.

Remotely sensed precipitation estimates derived from satellite platforms (Levizzani and Cattani, 2019) or radar observations (Bhusal et al., 2024) offer the advantage of extensive spatial coverage and high-spatial and high-temporal resolution monitoring, which are often unattainable with conventional ground-based networks. However, these estimates are susceptible to errors and systematic bias arising from retrieval algorithms, sensor limitations, and complex surface or atmospheric conditions (Wang et al., 2021). As a result, ground-based observations from rain gauges remain indispensable for the calibration, validation, and bias correction of remotely sensed precipitation products (Chen et al., 2019). This is particularly critical in rural and remote regions, where sparse gauge networks introduce significant uncertainties into satellite- or radar-derived rainfall estimates. Integrating high-quality in situ measurements with remotely sensed estimates is therefore essential for producing reliable precipitation datasets that support hydrological modeling, water resource management, and climate studies.

The growth of the Internet of Things (IoT) has transformed environmental monitoring by enabling distributed and low-cost sensor networks capable of transmitting data in real-time. These systems integrate sensor nodes, microcontrollers, and wireless communication modules (Anagnostou et al., 2004; Al-Hafiz et al., 2024), with time-stamped data transmitted to cloud platforms for real-time visualization, storage, and quality control (Singh et al., 2022; Ogunbunmi et al., 2024). Advances in energy-efficient design, including solar-powered operation and duty-cycling strategies, allow long-term autonomous deployment in remote environments with minimal power demand (Lai et al., 2016; Shen et al., 2023) for continuous and scalable environmental observation (He et al., 2024; Wang et al., 2020). The ability to transmit data instantly via LTE, LoRa, or Wi-Fi networks facilitates rapid access and integration with online visualization platforms. Collectively, these capabilities make IoT-based monitoring systems well-suited for capturing both short-term storm dynamics and long-term rainfall patterns.

In this study, we introduce the Low-Cost Efficient Wireless Intelligent Sensor (LEWIS 6), a compact and scalable rainfall monitoring platform. LEWIS 6 integrates a high-precision rainfall sensor with a Particle Boron LTE microcontroller, enabling accurate data acquisition and wireless transmission through LTE-M or NB-IoT networks. The system is powered by a small solar panel and a rechargeable lithium-ion battery, ensuring continuous operation even in remote or off-grid locations. Its design prioritizes low power consumption, operational reliability, and deployment flexibility, making it suitable for large-scale and long-term environmental monitoring applications. Furthermore, the modular architecture of LEWIS 6 allows easy integration of additional sensors or communication modules, facilitating customization for diverse hydrological and meteorological monitoring needs.

Field tests conducted in the State of New Mexico demonstrate that LEWIS 6 can provide high-spatial resolution, real-time precipitation data comparable to conventional measurement systems, but at a lower cost and with minimal maintenance requirements. The system's simplicity, energy efficiency, and robust connectivity make it particularly well-suited for widespread environmental monitoring, especially in regions lacking dense gauge networks. These findings highlight the potential of LEWIS 6 as a low-cost, reliable solution for improving hydrological data coverage and supporting climate resilience in data-scarce environments. The affordability and modular design of the LEWIS 6 platform enable the establishment of dense monitoring networks, which can improve hydrological model calibration, flood forecasting, and water resource management (Arante et al., 2025; Dhebe et al., 2023). Moreover, real-time precipitation data generated by LEWIS 6 support emergency response and agricultural decision-making by enhancing situational awareness during extreme rainfall events and optimizing irrigation practices (Peeriga et al., 2024; Raman and Iqbal, 2024). IoT-based rainfall monitoring systems such as LEWIS 6 are therefore critical for bridging the gap between sparse ground observations and large-scale satellite precipitation estimates, providing accurate, site-specific precipitation data essential for climate-resilient environmental monitoring.

## 2. Methodology

### 2.1 Sensor Hardware Configuration

The LEWIS 6 platform was designed and developed to operate as an energy-autonomous and self-sufficient network capable of continuous rainfall monitoring. The platform is specifically engineered to function independently in filed conditions, minimizing the need for external power sources and manual intervention. Each LEWIS 6 node integrates a set of selected hardware components to achieve reliable performance and energy efficiency (Figure 1). The core components of a LEWIS 6 node include:(a) solar panel, which provides renewable energy to power the system; (b) 1.5A USB/DC solar lithium-ion charger, responsible for regulating and optimizing the energy harvested from the solar panel; (c) lithium-ion battery, which stores the harvested energy to enable continuous operation during periods of low solar irradiance; (d) Particle Boron LTE microcontroller, serving as the computational and communication hub, capable of collecting sensor data and transmitting it via cellular LTE networks; (e) RJ11 analog adapter module, which facilitates the interface between the microcontroller and the rainfall sensing unit; and (f) rain collector unit, designed to measure and capture precipitation events with high-temporal resolution.

Power is supplied to the Particle Boron LTE microcontroller, which functions as the central control unit of the system. The microcontroller orchestrates a range of critical operations, including real-time data acquisition from connected sensors, signal processing to extract relevant features, and temporary local storage to ensure data integrity and continuity. In addition, the Boron LTE facilitates wireless communication, enabling seamless transmission of processed data to a cloud-based database for further analysis, long-term storage, and remote monitoring. Its integrated LTE connectivity ensures reliable and low-latency data transfer, making it particularly suitable for distributed or remote applications where consistent network access may be limited. By consolidating sensing, computation, and communication capabilities within a single platform, the Particle Boron LTE microcontroller significantly simplifies system architecture and enhances overall operational efficiency.

A comprehensive depiction of the LEWIS 6 platform system architecture is presented in Figure 1, highlighting the integration of energy harvesting, data acquisition, and communication subsystems that collectively facilitate autonomous rainfall monitoring in remote or off-grid locations. The architecture is designed to operate with minimal human intervention, leveraging renewable energy sources to sustain continuous operation even in resource-constrained environments. The modular configuration of the platform enables each sensor node to function independently while maintaining interoperability within a broader network, thereby supporting scalable deployment across extensive hydro-meteorological monitoring infrastructures. This design approach not only enhances system reliability and reduces maintenance requirements but also permits flexible adaptation to diverse environmental conditions and site-specific constraints, ensuring consistent data collection and long-term operational sustainability.

The system is engineered to efficiently manage both communication and power distribution across all connected components, ensuring uninterrupted operation under diverse and often harsh field conditions. Precipitation is measured using a high-precision tipping bucket rain sensor, which converts rainfall events into a sequence of electrical pulses proportional to the accumulated precipitation volume. To maintain signal stability and protect the microcontroller from electrical transients, the sensor is interfaced through an RJ11 adapter board. This interface not only ensures reliable electrical connectivity but also simplifies sensor replacement and maintenance. The acquired data are processed locally within the microcontroller, accurately time-stamped, and transmitted to a cloud-based platform in real-time, allowing for immediate visualization, automated analysis, and long-term archival of rainfall records.

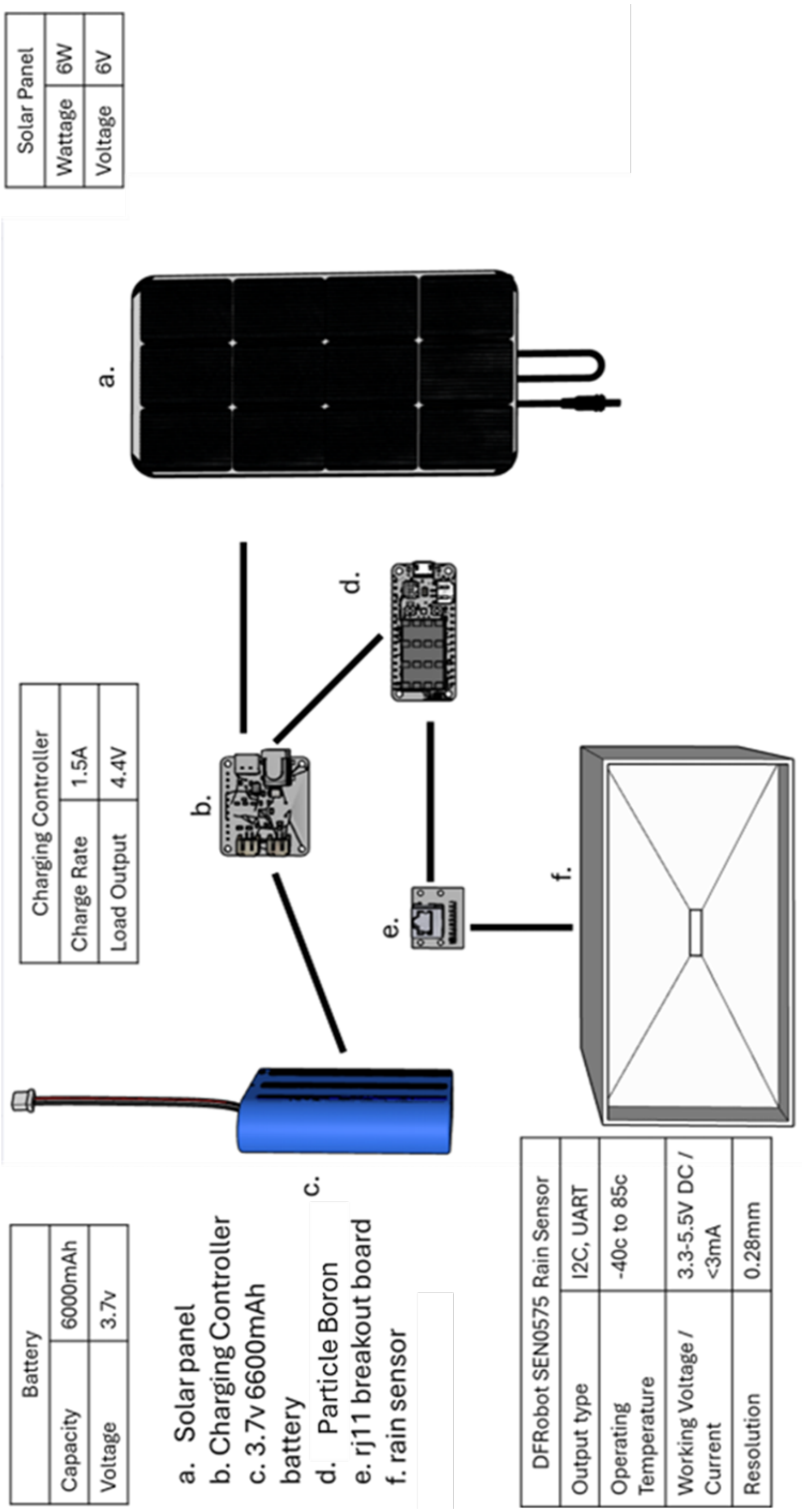

**Figure 1. Block diagram of the solar-powered LEWIS 6 rainfall monitoring system: (a) Solar panel supplies renewable energy; (b) the charge controller regulates power flow to (c) the lithium-ion battery used for energy storage; (d) the Particle Boron LTE board manages data processing and wireless transmission; (e) RJ11 board; and (f) the tipping-bucket rain sensor provide real-time precipitation measurements, supported by an external antenna for reliable connectivity.**

The core of the rain monitoring system comprises two successive generational architectures, LEWIS 5 and LEWIS 6, both designed to deliver real-time precipitation measurements with minimal cost and power consumption. These architectures incorporate modular design principles, allowing individual sensor nodes to be easily integrated, replaced, or expanded within larger monitoring networks. Such modularity enhances system scalability, making the platform suitable for dense sensor deployments across wide geographic areas. Calibration procedures are implemented to ensure consistent measurement accuracy across all nodes, accounting for local environmental variations such as temperature and wind effects.

In addition to hardware considerations, the system architecture emphasizes data reliability and integrity. Local data buffering allows for temporary storage during network interruptions, preventing loss of critical measurements. Furthermore, time-synchronized data logging facilitates integration with other meteorological datasets, enabling comprehensive environmental analyses. Collectively, these design features ensure that the LEWIS 6 platform provides high-fidelity, continuous, and autonomous rainfall monitoring with minimal operational maintenance, offering a robust solution for large-scale hydro-meteorological studies and environmental monitoring initiatives.

LEWIS 5 is characterized by a modular design that employs the Arduino Uno Rev3 microcontroller as its central processing unit. Precipitation measurements are obtained using a YL-83 sensor, which is interfaced through the microcontroller's analog input. Data transmission to the cloud is facilitated via an ESP8266 Wi-Fi. The system's energy requirements are met through a Verizon Jetpack hotspot and a DC-DC converter (XL6009), powered continuously by a battery. This architecture, however, is inherently energy-intensive, with each component drawing significant power. Consequently, LEWIS 5 is dependent on external, non-renewable power sources or frequent battery replacements, which limits its suitability for extended or remote deployments.

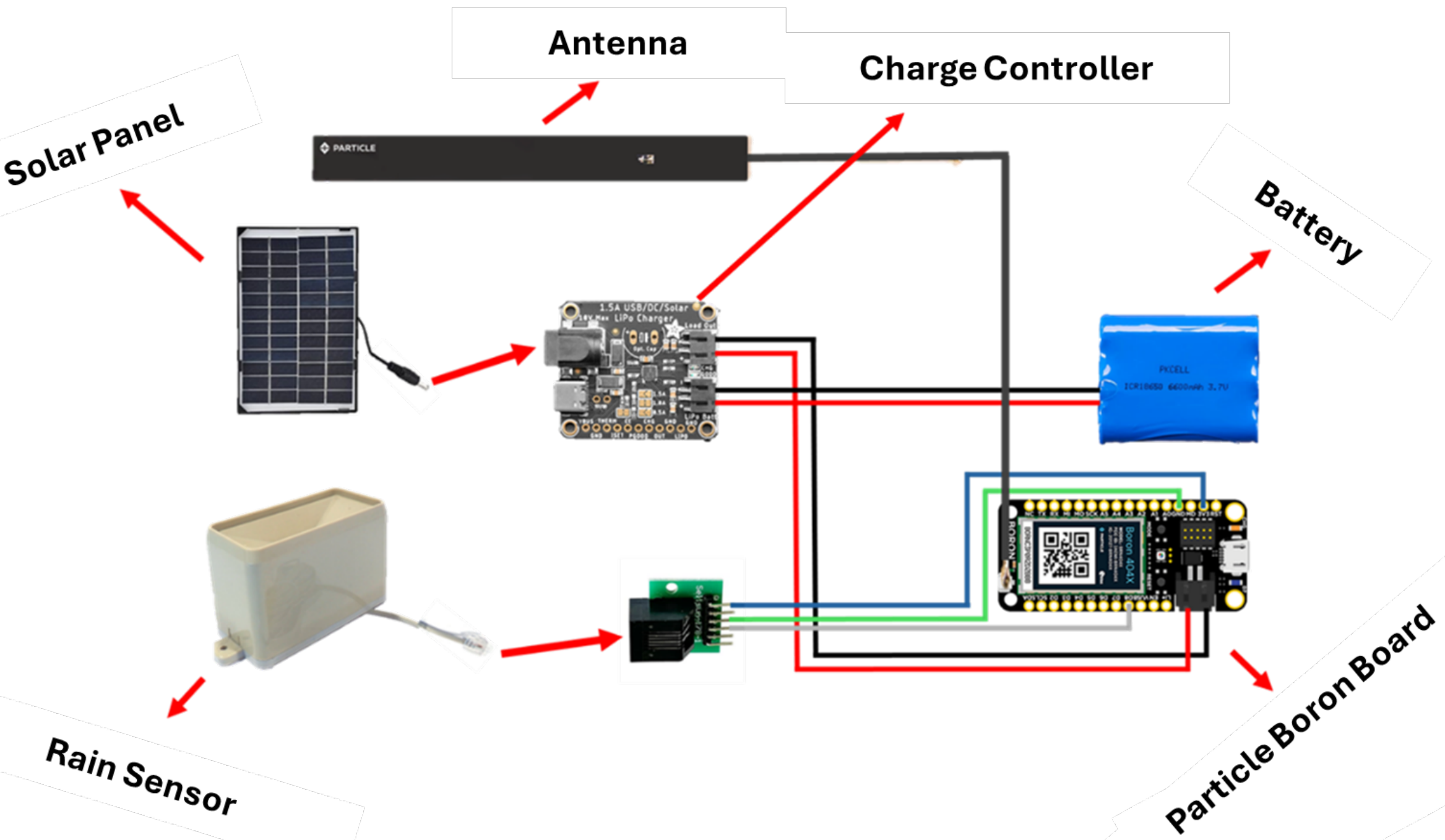

**Figure 2. A schematic diagram of the LEWIS 6 rainfall monitoring node; it shows the integration of the rain sensor, Particle Boron LTE board, charge controller, battery pack, solar panel, and antenna for continuous, solar-powered data acquisition and wireless transmission.**

In contrast, the LEWIS 6 platform (schematically illustrated in Figure 2) is specifically optimized for energy efficiency and solar-powered autonomy. This design replaces the Arduino and Wi-Fi module with a fully integrated and low-power Particle Boron microcontroller, which supports LTE-M/NB-IoT connectivity for reliable long-range data transmission. The power subsystem incorporates a 3.7V, 35Ah lithium-ion battery, a solar charge controller, and a 25W, 12V Thunderbolt Magnum solar panel, enabling autonomous recharging. The charger circuit is based on Depvko 20A controllers, providing a continuous current rating sufficient to support both sensor operation and periodic data transmission duty cycling.

The sensors are housed within a weatherproof enclosure, precisely aligned to ensure accurate rainfall collection under field conditions. This hardware optimization, combined with the energy-efficient architecture, enables long-term, maintenance-free operation, even in remote and off-grid locations. Collectively, these enhancements represent a significant advancement over the LEWIS 5 generation, improving system autonomy, reliability, and scalability for large-scale environmental monitoring applications. The sensors are housed within a weatherproof casing, with precise alignment for rainfall collection. The hardware improvements enable long-term maintenance-free deployments in remote areas and make LEWIS 6 a major advancement over LEWIS 5.

## 2.2. Energy Profile Comparison between LEWIS 5 and LEWIS 6

The transition from LEWIS 5 to LEWIS 6 represents one of the most significant improvements in the evolution of the LEWIS platform. LEWIS 5, while effective as a proof of concept, consumed approximately 144 Wh/day due to its reliance on Arduino Uno with an ESP8266 Wi-Fi module (Table 1). Continuous Wi-Fi connectivity and the absence of deep sleep functionality made the system highly energy-intensive, necessitating larger power systems, including 12V sealed lead-acid batteries and larger solar panels. This configuration reduced portability and reliability, particularly in remote environments with inconsistent solar availability. Although field deployment was feasible, it was not sustainable for long-term or large-scale monitoring networks due to frequent maintenance requirements and high power consumption.

**Table 1. LEWIS 5 energy profile.**

| Component | Model | Power (W) | Active Time (h/day) | Energy Usage (Wh/day) |
|---|---|---|---|---|
| Arduino Uno | Rev3 | 0.5 | 24 | 12 |
| Rain Sensor | YL-83 | 0.25 | 6 | 6 |
| WiFi Shield | ESP8266 | 0.5 | 24 | 12 |
| DC-DC Converter | XL6009 | 2.0 | 24 | 48 |
| Hotspot | Verizon Jetpack | 2.75 | 24 | 66 |
| Total | | | | 144 |

LEWIS 6 introduced a paradigm shift through the integration of the Particle Boron microcontroller, which natively supports low-power LTE Cat-M1 communication. This transition significantly reduced transmission energy costs, as LTE-M is specifically optimized for intermittent and low-bandwidth IoT data transfer with minimal overhead. In contrast to the continuous connectivity requirements of earlier designs, this architecture enables short, yet efficient communication bursts followed by extended low-power states. As a result, the system can minimize radio-on time while maintaining reliable data delivery, which is particularly important for remote environmental deployments where energy availability is limited, and transmission efficiency directly impacts system longevity.

In addition, LEWIS 6 incorporates advanced power management strategies, including deep sleep modes, adaptive duty cycling, and highly efficient charging circuits with maximum power point tracking (MPPT). Together, these enhancements substantially reduce overall energy demand, lowering daily energy consumption to approximately 7 Wh/day (Table 2), nearly 20 times lower than LEWIS 5. As a result, the system can operate reliably using a compact 1 W solar panel and a small 3.7V lithium-ion battery, thereby eliminating the bulk, weight, and cost associated with earlier power designs. This dramatic increase in energy efficiency enhances the autonomy of individual sensor nodes and enables scalable deployment across distributed field sites. LEWIS 6 can therefore be deployed in larger networks without the logistical burden of maintaining oversized solar arrays and heavy battery packs.

**Table 2. LEWIS 6 Energy Profile.**

| Component | Voltage (V) | Current (mA) | Power (W) | Hours Active | Energy Usage (Wh/day) |
|---|---|---|---|---|---|
| Microcontroller | 3.7 | 35 | 0.129 | 24 | 3.108 |
| Rain Sensor | 3.3 | 5 | 0.0165 | 6 | 0.099 |
| Adapter Module | 3.3 | 2 | 0.0066 | 24 | 0.158 |
| Charger Module | 3.7 | 5 | 0.0185 | 24 | 0.444 |
| Tx Energy | - | - | - | - | 2.9 |
| TOTAL | - | - | - | - | 6.709 |

Overall, the transition from LEWIS 5 to LEWIS 6 demonstrates how the integration of IoT-optimized hardware, efficient communication protocols, and compact energy harvesting systems can significantly improve the performance and sustainability of environmental sensing platforms. The reduction in power consumption, combined with improved system autonomy and reduced maintenance requirements, enables reliable long-term deployment in remote environments. These advancements position LEWIS 6 as a low-cost and highly reliable solution for real-time rainfall monitoring and provide a scalable framework for future distributed environmental monitoring networks.

## 2.3 Sensor Software

The software architecture of the rain sensor system is centered on real-time data acquisition, local data preprocessing, low-power operation, and efficient wireless transmission to a centralized cloud database. The system firmware is designed to ensure reliable environmental monitoring while minimizing energy consumption, particularly for long-term field deployment in remote locations.

For LEWIS 5, the firmware was developed in the Arduino IDE using C/C++. The program continuously samples analog signals generated by the YL-83 sensor and converts these raw readings into corresponding voltage or resistance values. Calibration algorithms are then applied to transform the sensor output into estimated rainfall depth measurements. The processed data are transmitted periodically to cloud-based platforms such as Google Firebase or ThingSpeak using either the HTTP or MQTT communication protocol. The software architecture follows a simple polling-based acquisition model in which sensor readings are collected continuously and uploaded at predefined intervals. While this implementation provides stable real-time monitoring and straightforward cloud integration, it exhibits relatively high power consumption because the Wi-Fi module requires continuous hotspot connectivity during operation. Consequently, the communication subsystem becomes the dominant source of energy usage, limiting battery life and reducing the suitability of the system for prolonged autonomous deployment.

In contrast, LEWIS 6 adopts a substantially more energy-efficient software architecture by leveraging the Particle Device OS and the native low-power capabilities of the Particle Boron LTE platform. Firmware development was performed using Particle Workbench integrated with Visual Studio Code. The system implements an event-driven and energy-aware operational strategy through the use of Particle's System.sleep() and System.on() functions. Instead of maintaining continuous operation, the device enters deep sleep mode for the majority of the duty cycle and wakes only at predefined intervals (e.g., every 10 minutes) to perform sensing and communication tasks. Upon waking, the microcontroller initializes the rain sensor, acquires measurement data, formats the readings into JSON structures, and transmits the information securely through the Particle Cloud API over LTE connectivity.

This duty-cycled approach substantially reduces overall energy consumption because both the microcontroller and associated peripherals remain in an ultra-low-power state when inactive. As a result, LEWIS 6 achieves significantly longer operational lifetime and improved battery efficiency compared with the Wi-Fi-based architecture of LEWIS 5. In addition to rainfall measurements, the firmware also records diagnostic parameters including battery voltage, LTE signal strength, connection status, and system error codes. These diagnostic data support remote health monitoring and facilitate predictive maintenance of deployed units.

In addition, the software architecture of LEWIS 6 supports over-the-air (OTA) firmware updates, enabling remote reconfiguration, debugging, and feature enhancement without requiring physical access to the device. The modular organization of the firmware allows straightforward scalability and feature expansion, including integration of multiple environmental sensors, GPS-based geotagging, adaptive sampling algorithms, and distributed rainfall mapping applications. This modular and low-power architecture therefore provides a flexible foundation for future IoT-based environmental monitoring systems.

### 2.4 Data Acquisition and Transmission Architecture

Sensor nodes are programmed to continuously sample rainfall data, with the Particle Boron microcontroller processing the raw analog signals acquired from the rain gauge through the RJ11 adapter module. The incoming analog measurements are converted into calibrated digital precipitation values using onboard signal conditioning and data acquisition routine implemented within the embedded firmware. Each observation is assigned a precise timestamp and encoded into a structured JSON packet containing precipitation depth, device identification number, timestamp, and battery voltage. This standardization data structure facilitates seamless wireless transmission, supports efficient parsing and storage, and ensures reliable cloud-based archiving of field measurements for subsequent analysis and long-term monitoring applications.

Data packets are transmitted in real-time via LTE-M communication using the MQTT protocol to the Particle cloud platform, as illustrated in Figure 3. The cloud infrastructure automatically receives and routes incoming telemetry messages to a custom-built web dashboard, where sensor data are visualized, monitored, and archived continuously. The dashboard provides a centralized interface for observing rainfall dynamics, evaluating sensor performance, and maintaining historical records of precipitation measurements. This cloud-connected architecture enables low-latency data delivery, remote accessibility, and scalable deployment of distributed rainfall monitoring networks across geographically diverse field locations.

Rain Sensor
Microcontroller (Boron)
LTE Protocol
Data to cloud
Cloud Storage
Our server
Lab Website

**Figure 3. Data flow architecture of the IoT-based rainfall monitoring system.**

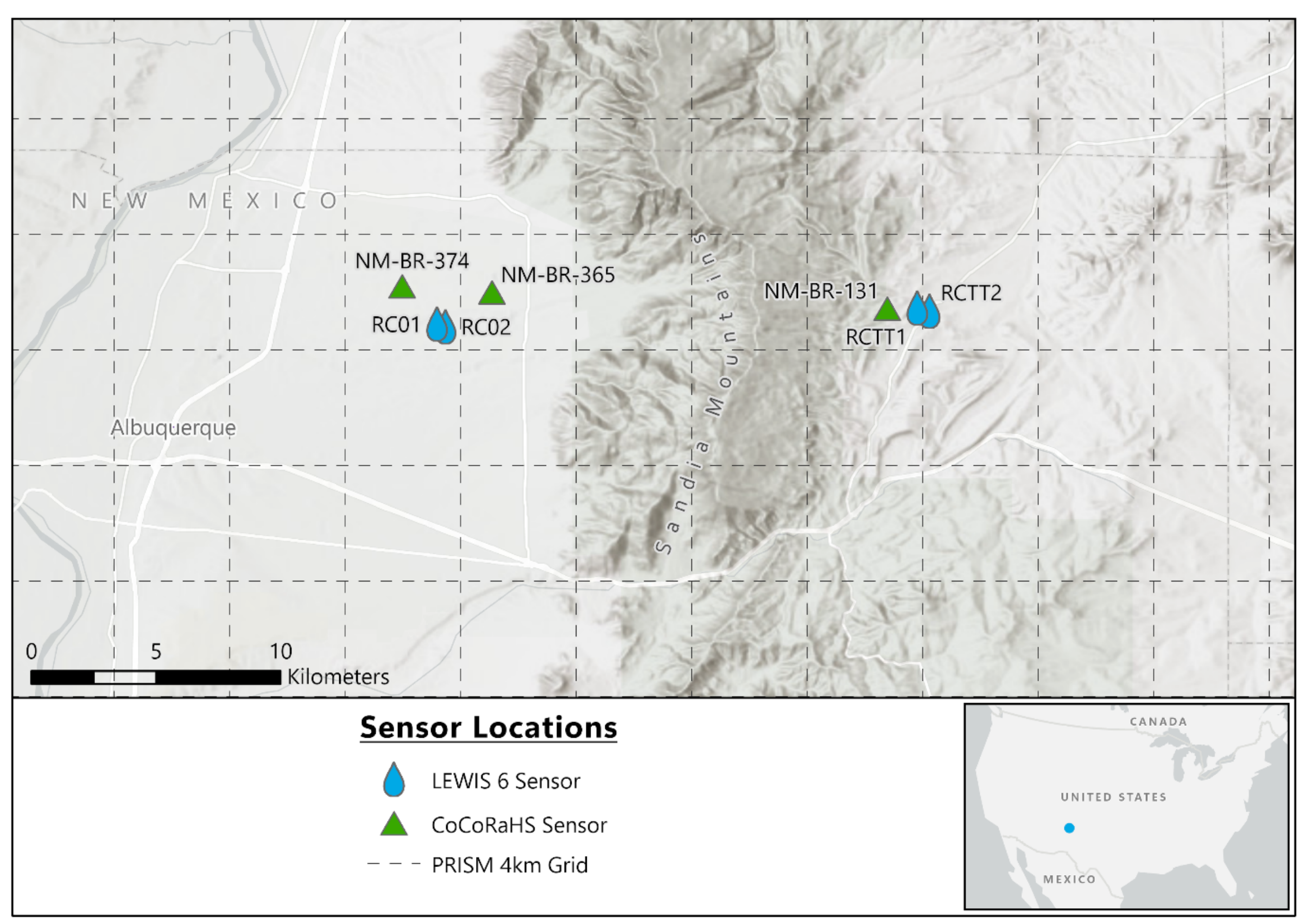

**Figure 4. Field deployment locations of LEWIS 6 sensors.**

## 2.5. Sensor Deployment

The map presented in Figure 4 shows the current deployment locations of the LEWIS 6 rainfall sensing network across central New Mexico. The selected deployment sites include RCTT1 and RCTT2 (Figure 5), situated in Tinkertown, as well as RC01 and RC02 (Figure 6), located in northeast Albuquerque. These sites were strategically selected to capture variability in local precipitation patterns while also ensuring site accessibility, stable sensor operation, and the availability of reliable reference datasets for validation purposes. The geographic distribution of the deployment locations enabled the evaluation of sensor performance under differing environmental and microclimatic conditions of central New Mexico.

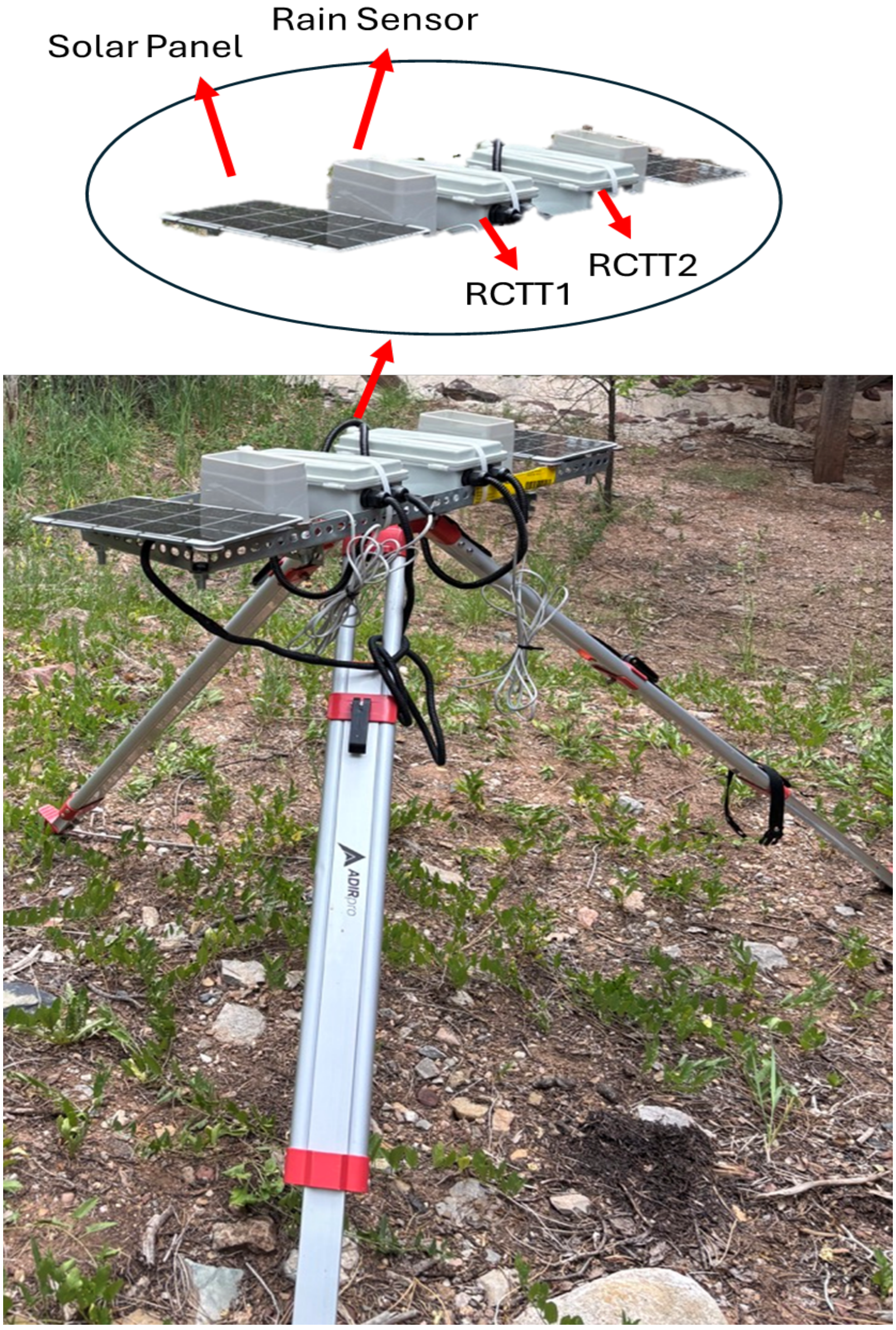

**Figure 5: Field deployment of a RCTT1 and RCTT2 rainfall sensor node.**

At each deployment site, a LEWIS 6 node continuously monitors precipitation and transmits time-stamped observations to a centralized cloud-based database for storage, processing, and subsequent analysis. The deployment architecture is designed to support uninterrupted environmental monitoring and near real-time data accessibility, thereby enabling efficient assessment of rainfall events and sensor performance over extended observation periods. Each sensor node operates autonomously in the field, recording precipitation depth measurements together with supporting system diagnostics, including device identification and power status, to ensure data traceability and operational reliability.

To evaluate measurement quality and system consistency, co-located LEWIS 6 sensors were initially validated against one other to assess internal agreement, repeatability, and short-term measurement stability under identical environmental conditions. This inter-sensor comparison provided an important first-stage evaluation of the reproducibility of the sensing platform and helped identify potential systematic biases or calibration inconsistencies between units. Following the internal validation phase, LEWIS 6 observations were compared against independent precipitation datasets obtained from established reference sources. These comparisons enable assessment of the accuracy and reliability of the LEWIS 6 network relative to conventional rainfall monitoring systems and publicly available precipitation records.

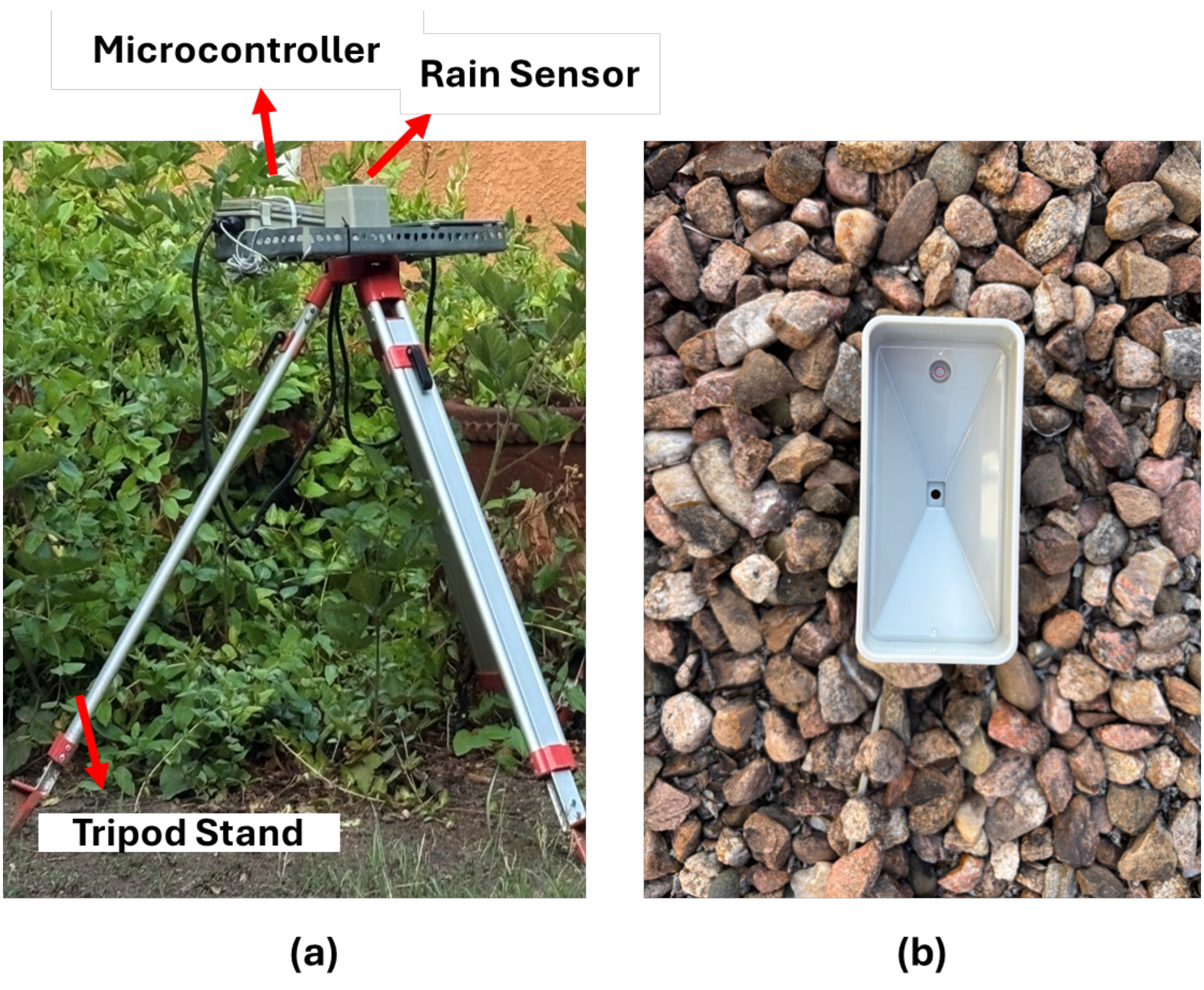

**Figure 6. Field deployment of RC01 (a) and a close view of RC02 rainfall sensor node (b).**

## 2.6. Validation Strategy

Ground-based reference observations were obtained from the Community Collaborative Rain, Hail, and Snow Network (CoCoRaHS), a nationwide volunteer monitoring program that provides point-scale precipitation measurements using standardized manual gauges with 0.01 inch (~0.25 mm) resolution (https://www.cocorahs.org/). Because CoCoRaHS observations represent localized precipitation accumulation at specific measurement sites, they serve as valuable ground-truth data for validating automated sensor observations and for capturing fine-scale spatial variability associated with convective and orographic precipitation processes.

Model-based precipitation estimates were obtained from the Parameter-elevation Regressions on Independent Slopes Model (PRISM), a widely used gridded climatological dataset developed at Oregon State University. PRISM generates spatially continuous precipitation fields by integrating station observations with regression-based climate-elevation relationships and physiographic modifiers, including elevation, slope aspect, coastal proximity, and terrain shadowing (Daly et al., 2008). The resulting gridded dataset, available at approximately 4-km spatial resolution, provides consistent regional-scale precipitation estimates that are well-suited for watershed-scale hydrologic and climate applications (Jeton, 2005; Marraccini et al., 2014). Daily precipitation recorded by each LEWIS sensor was compared against PRISM estimates extracted for the corresponding geographic coordinates and time periods. While PRISM provides robust regional coverage, its spatial interpolation and smoothing procedures may reduce sensitivity to localized precipitation maxima and short-duration storm events, particularly in complex terrain (Hwang and Ham, 2013). This limitation further motivates the incorporation of in situ sensor validation.

In addition, radar-based quantitative precipitation estimates derived from Next Generation Weather Radar (NEXRAD) system and obtained from the National Centers for Environmental Information (https://www.ncei.noaa.gov/products/radar/next-generation-weather-radar) were used for event-scale evaluation. NEXRAD Level III products estimate accumulated precipitation using reflectivity-rainfall relationships and provide temporally resolved precipitation fields that are particularly useful for examining storm evolution and short-duration rainfall dynamics. Unlike point-based gauge measurements or daily gridded climate products, radar observations offer near-continuous spatial and temporal coverage across the study domain for selected storm events, thereby enabling detailed sub-daily comparisons with LEWIS 6 precipitation measurements. The spatial continuity of radar estimates also allows for evaluation of localized precipitation variability that may not be fully represented by sparsely distributed ground-based observing networks. Although radar-derived precipitation products are subject to uncertainties associated with beam geometry, attenuation, bright-band contamination, and assumptions embedded within reflectivity-rainfall conversion algorithms, they remain highly effective for characterizing storm-scale precipitation processes and temporal rainfall evolution.

Together, the combined use of regional gridded datasets from PRISM, community-based ground observations from CoCoRaHS, and radar-derived precipitation estimates from NEXRAD provides a complementary validation framework for precipitation analysis. This integrated approach enables the assessment of both broad-scale spatial precipitation patterns and fine-scale local rainfall variability, thereby improving the robustness and reliability of precipitation evaluation across multiple temporal and spatial scales. By combining independently derived observational products, the framework also helps reduce uncertainty associated with any single measurement source, while allowing for cross-comparison between gridded climatological datasets, in situ observations, and remotely sensed radar estimates.

### 2.7 Performance Assessment

To evaluate agreement among datasets, the root mean square error (RMSE) was selected as the primary performance metric because it provides a direct measure of the magnitude of differences between observed and reference precipitation values. RMSE quantifies overall prediction error while retaining the physical units of rainfall, thereby allowing straightforward interpretation in terms of rainfall measurement accuracy. This metric is particularly well suited for precipitation comparisons because it places greater emphasis on larger deviations, making it especially sensitive to discrepancies associated with high-intensity storm events hydrologically significant. Given that the primary objective of this study was to assess measurement accuracy rather than variability alone, RMSE provides a more meaningful indicator of dataset agreement across multiple temporal scales. In addition, its consistent formulation facilitates direct comparison among datasets and aggregation periods. Therefore, RMSE was adopted as the standardized evaluation metric for monthly, weekly, and daily precipitation comparisons throughout the study. The RMSE is defined as follows:

$$RMSE = \sqrt{\left\{\left(\frac{1}{N}\right)\sum_{i=1}^{N}\left(x_i - y_i\right)^2\right\}} \qquad (1)$$

where $x_i$ and $y_i$ represent paired precipitation observations from the two datasets being compared at time step $i$ (e.g., LEWIS 6 sensor measurements, CoCoRaHS gauge observations, or PRISM model estimates), and N denotes the total number of observations included in the comparison.

## 3. Results and Discussion

### 3.1 Monthly Precipitation Performance

3.1.1 RCTT1 and RCTT2 vs PRISM vs CoCoRaHS gauge

Monthly comparison in Figure 7 (a) shows a comparison of co-located LEWIS 6 sensors (RCTT1 and RCTT2), PRISM model data, and the nearby CoCoRaHS gauge data for August and September 2025. All datasets captured the seasonal increase in precipitation from August to September, although differences in magnitude were observed among the measurement sources. CoCoRaHS consistently reported the highest monthly precipitation totals, followed by the two field sensors, whereas PRISM generally produced slightly lower values. This reduction in PRISM precipitation totals likely reflects the spatial smoothing inherent to its 4-km grid resolution, which represents regional-scale precipitation patterns rather than localized point measurements.

The two co-located LEWIS 6 sensors exhibited the lowest error (RMSE = 0.24 in), indicating a high degree of measurement consistency and repeatability at the same monitoring site. Comparisons between the sensor observations and PRISM estimates produced moderately higher RMSE values ranging from 0.34 to 0.58 in, suggesting systematic differences between point-scale observations and gridded precipitation estimates. When evaluated against the independent CoCoRaHS gauge observations, RCTT2 demonstrated the strongest agreement (RMSE = 0.24 in), followed by RCTT1 (RMSE = 0.48 in), whereas PRISM exhibited the largest discrepancy (RMSE = 0.81 in).

Overall, the RMSE indicates that the LEWIS 6 sensors effectively capture localized rainfall magnitudes with strong observational consistency, while PRISM more accurately represents broader regional precipitation trends but smooths fine-scale, site-specific precipitation variability. In this comparison framework, CoCoRaHS served as an effective ground-based reference dataset for evaluating precipitation measurement accuracy.

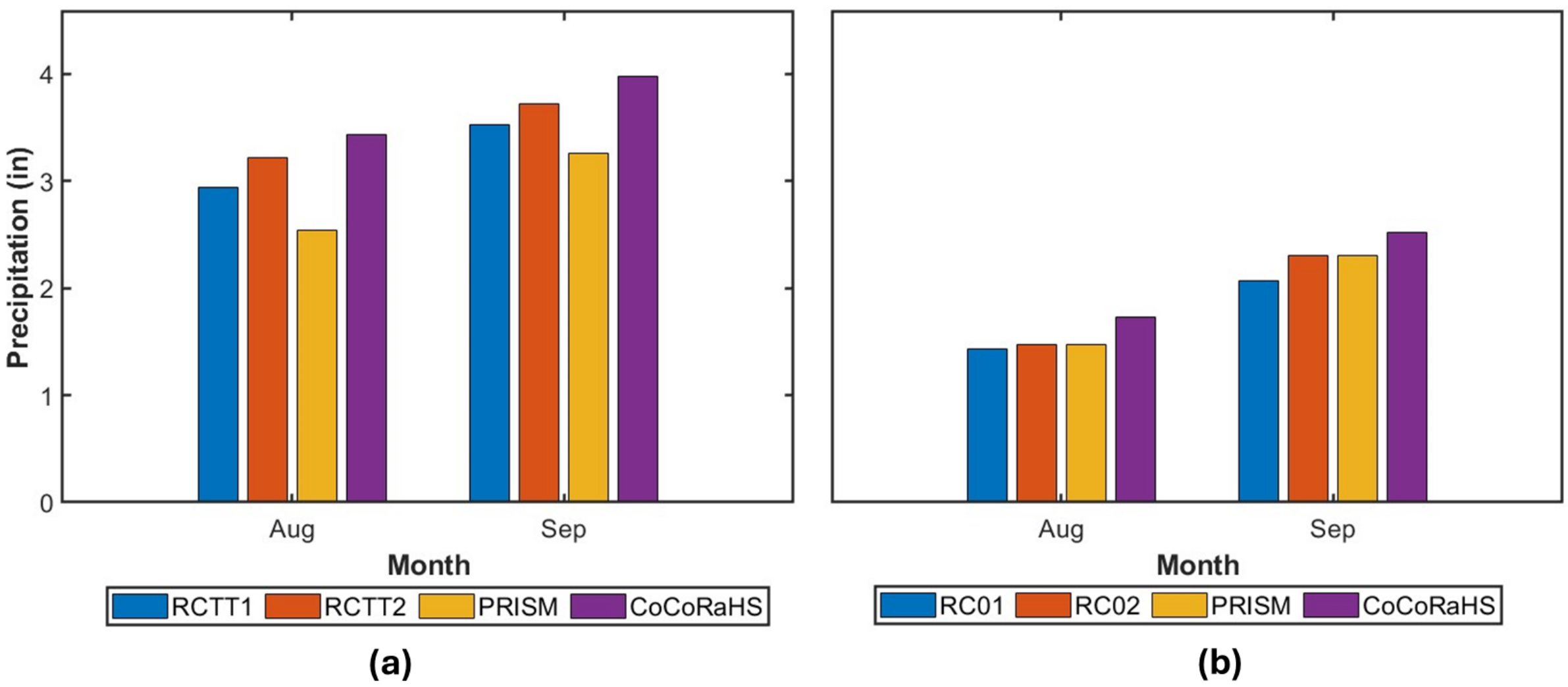

**Figure 7: Monthly precipitation comparison for August and September 2025 between LEWIS-6 sensors, PRISM gridded precipitation data, and nearby CoCoRaHS reference gauges. (a) Tinkertown site showing RCTT1, RCTT2, PRISM, and CoCoRaHS totals. (b) Northeast Albuquerque site showing RC01, RC02, PRISM, and CoCoRaHS totals.**

*3.1.2 RC01 and RC02 vs PRISM vs CoCoRaHS gauge*

A similar pattern was observed in Figure 7 (b) for the Northeast Albuquerque site, where RC01, RC02, PRISM, and CoCoRaHS all captured the seasonal increase in precipitation from August to September, 2025. The two LEWIS 6 sensors showed close agreement, while CoCoRaHS generally recorded slightly higher totals, suggesting localized rainfall intensification that may not be fully represented by either the LEWIS sensors or the PRSIM gridded model. PRISM closely followed the temporal behavior of the field instruments (LEWIS 6 and CoCoRAHS), indicating that the broader regional rainfall distribution was well represented at this location. The co-located LEWIS 6 sensors exhibited the lowest error (RMSE = 0.09 in), confirming stable and repeatable sensor performance. Comparisons with PRISM produced RMSE values of 0.34 in for the RC01–PRISM comparison and 0.23 in for the RC02–PRISM comparison, indicating modest differences between gridded model estimates and point-based ground observations. When evaluated against CoCoRaHS, RC02 demonstrated close agreement (RMSE = 0.24 in), whereas RC01 exhibited slightly higher error (RMSE = 0.38 in), suggesting the presence of minor localized precipitation variability or small differences in gauge response.

Overall, the RMSE analysis indicates that the LEWIS 6 sensors remained highly consistent with each other throughout the observation period, while PRISM reliably represented broader regional precipitation patterns at this study site, albeit with moderate smoothing relative to point-based observations.

## 3.2. Weekly Precipitation Performance

### *3.2.1. RCTT1 and RCTT2 vs PRISM vs CoCoRaHS gauge*

Weekly precipitation comparison at the Tinkertown site demonstrated consistent temporal behavior among all datasets while revealing greater short-term variability relative to the monthly analysis (Figure 8 (a)). All datasets reflected the same sequence of rainfall events, including low accumulation during early August, moderate mid-month precipitation, and a pronounced rainfall peak during late September 2025. The two co-located LEWIS 6 sensors (RCTT1 and RCTT2) closely tracked each other throughout the nine-week period, whereas PRISM and CoCoRaHS occasionally exhibited larger deviations during individual storm weeks, reflecting spatial heterogeneity and localized differences in rainfall intensity. The co-located sensors exhibited the lowest error (RMSE = 0.08 in), confirming stable and repeatable performance at the same location. Comparisons with PRISM produced substantially higher RMSE values (0.35–0.36 in), indicating that the gridded model smooths short-duration rainfall events and does not fully capture localized weekly peaks. When evaluated against the independent CoCoRaHS gauge, RCTT2 showed the closest agreement (RMSE = 0.15 in), followed by RCTT1 (RMSE = 0.18 in), whereas PRISM exhibited larger discrepancies (RMSE = 0.30 in). Overall, the RMSE results indicate that the LEWIS 6 sensors provided accurate and locally representative short-term precipitation measurements, whereas PRISM was more effective at representing broader regional precipitation trends than event-scale rainfall variability.

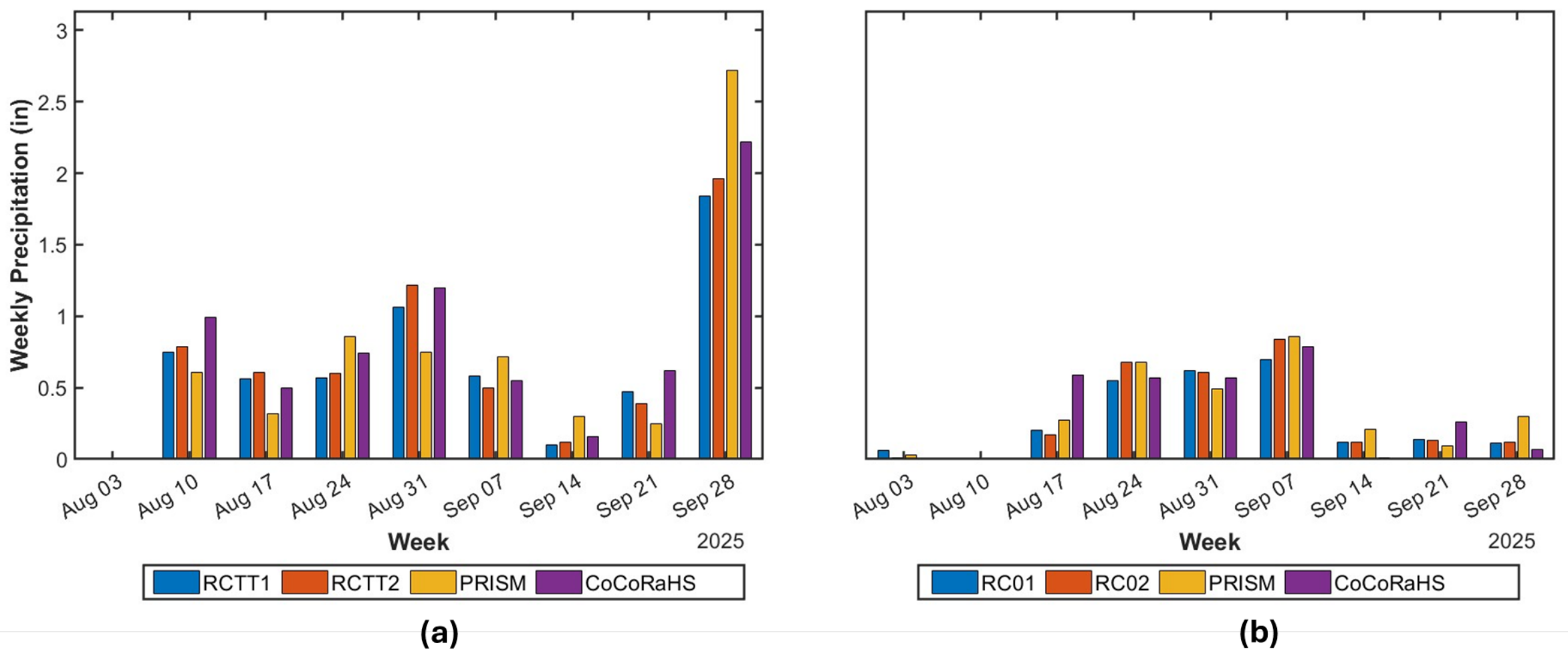

**Figure 8. Weekly precipitation totals for August-September 2025 comparing LEWIS-6 sensors, PRISM gridded precipitation data, and nearby CoCoRaHS reference gauges at two locations. (a) Tinkertown site showing RCTT1, RCTT2, PRISM, and CoCoRaHS weekly rainfall. (b) Northeast Albuquerque site showing RC01, RC02, PRISM, and CoCoRaHS weekly rainfall.**

### *3.2.2. RC01 and RC02 vs PRISM vs CoCoRaHS gauge*

A similar weekly pattern was observed at the Northeast Albuquerque site, where RC01, RC02, PRISM, and CoCoRaHS all reproduced the general timing of precipitation events but exhibited greater week-to-week fluctuations than seen in the monthly aggregates (Figure 8(b)). The two LEWIS 6 sensors closely followed each other across the measurement period, particularly during mid-September storms, whereas larger differences occasionally appeared between the point measurements and the PRISM model or community gauge during isolated events. These discrepancies highlighted the sensitivity of weekly rainfall to localized storm structure and micro-scale variability. The co-located sensors exhibited the lowest error (RMSE = 0.07 in), demonstrating strong internal consistency of the field instrumentation. Comparisons with PRISM produced RMSE values ranging from 0.09 to 0.11 in, suggesting that the model reasonably represents regional weekly rainfall patterns but still smooths short-duration peaks. In contrast, comparisons with CoCoRaHS showed higher RMSE values ranging from 0.15 to 0.17 in, indicating increased local variability and gauge-to-gauge differences at the weekly scale.

## 3.3 Daily Precipitation Performance

### *3.3.1. RCTT1 and RCTT2 vs PRISM vs CoCoRaHS gauge*

The daily precipitation comparison at the Tinkertown site revealed substantially greater short-term variability than observed at the weekly and monthly aggregation levels (Figure 9). All datasets captured the timing of individual storm events, including moderate rainfall in mid-August, scattered small events through early September, and a pronounced precipitation peak on September 28, 2025. The two co-located LEWIS 6 sensors (RCTT1 and RCTT2) tracked each other closely for nearly all days, while larger discrepancies were occasionally observed between the point measurements and the gridded PRISM model or the nearby CoCoRaHS gauge, particularly during high-intensity events. The co-located sensors exhibited the lowest error (RMSE = 0.04 in), confirming highly consistent and repeatable measurements at the same location. Comparisons with PRISM showed markedly larger RMSE values (0.28–0.29 in), indicating that the gridded model smooths localized rainfall peaks and underrepresents event-scale extremes. Relative to CoCoRaHS, both sensors outperformed the PRISM model, with RCTT2 demonstrating the closest agreement (RMSE = 0.19 in). Overall, the daily analysis confirms that the LEWIS 6 sensors provided more precise and locally representative event-scale precipitation measurements, whereas PRISM was better suited for capturing broader regional patterns. The increase in RMSE from monthly to weekly to daily scales further highlighted the sensitivity of short-duration rainfall to micro-scale spatial variability, emphasizing the importance of in-situ gauges for high-resolution hydrologic monitoring.

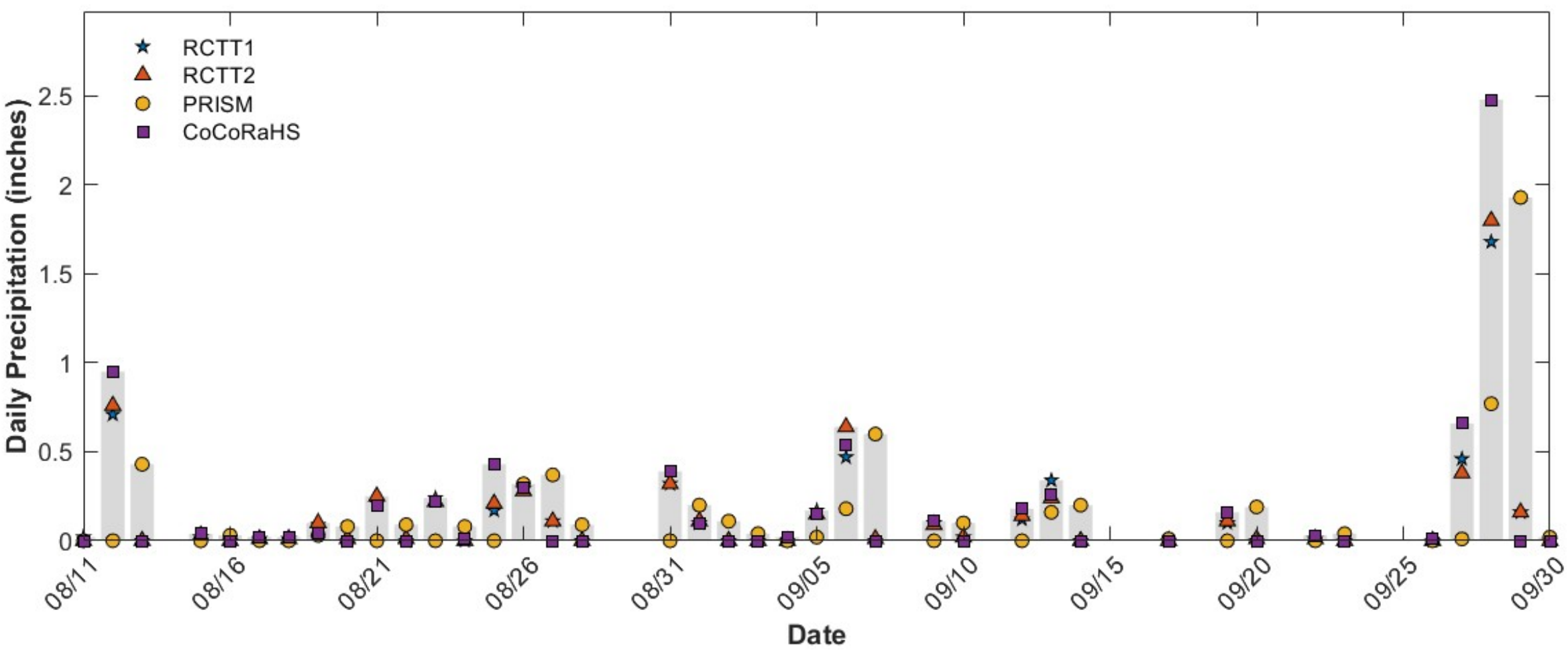

**Figure 9. Daily precipitation totals at the Tinkertown site comparing the co-located LEWIS-6 sensors (RCTT1 and RCTT2), PRISM gridded precipitation data, and the CoCoRaHS community reference gauge for August–September 2025.**

*3.3.2. RC01 and RC02 vs PRISM*

A similar pattern was observed at the Northeast Albuquerque site, where daily rainfall totals from RC01, RC02, PRISM, and CoCoRaHS consistently reproduced the occurrence of individual storm events but showed increased day-to-day variability compared with aggregated weekly or monthly totals (Figure 10). Gray bars represent the highest daily rainfall among all three datasets. The two LEWIS 6 sensors again followed nearly identical trends, particularly during late August and early September rainfall events, while larger differences were observed between the point measurements and both the PRISM model and CoCoRaHS gauge during isolated storms. These discrepancies reflected localized rainfall heterogeneity and the influence of small-scale convective activity that may not be fully resolved by regional or community-based observations. The co-located sensors exhibited the lowest error (RMSE = 0.03 in), confirming excellent internal consistency. Comparisons with PRISM produced RMSE values ranging from 0.11 to 0.12 in, indicating that the model captures general regional trends but smooths event-scale fluctuations. Higher RMSE values were observed relative to CoCoRaHS (0.17–0.19 in), reflecting increased gauge-to-gauge variability at daily resolution. Overall, these results demonstrated that while PRISM provides reasonable long-term regional estimates, LEWIS 6 sensors offered superior precision for capturing short-term rainfall dynamics. The consistently low RMSE between sensor pairs across all temporal scales reinforced the reliability of the deployed field instrumentation for high-frequency precipitation monitoring.

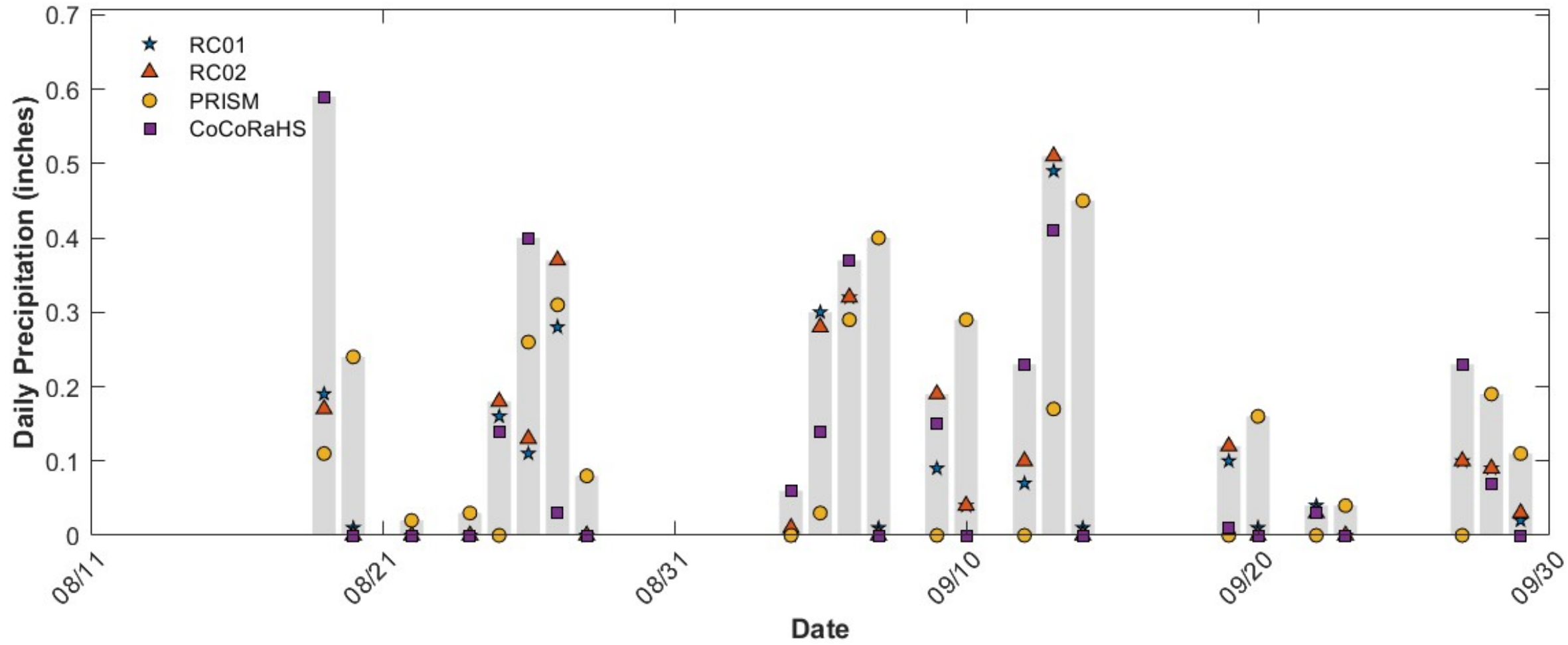

**Figure 10. Daily precipitation totals at the Northeast Albuquerque site comparing LEWIS-6 sensors (RC01 and RC02), PRISM gridded precipitation data, and the CoCoRaHS community reference gauge for August–September 2025.**

## 3.4 Pattern mismatch with PRISM

*3.4.1. RCTT1 and RCTT2 vs PRISM*

Figure 11 illustrates daily precipitation recorded by two LEWIS 6 sensors (RCTT1 and RCTT2) compared with PRISM model data from August to September 2025. Gray bars represent the highest daily rainfall among all three datasets. Red and blue horizontal lines indicate instances where the LEWIS 6 sensor readings and PRISM estimates followed different temporal rainfall patterns.

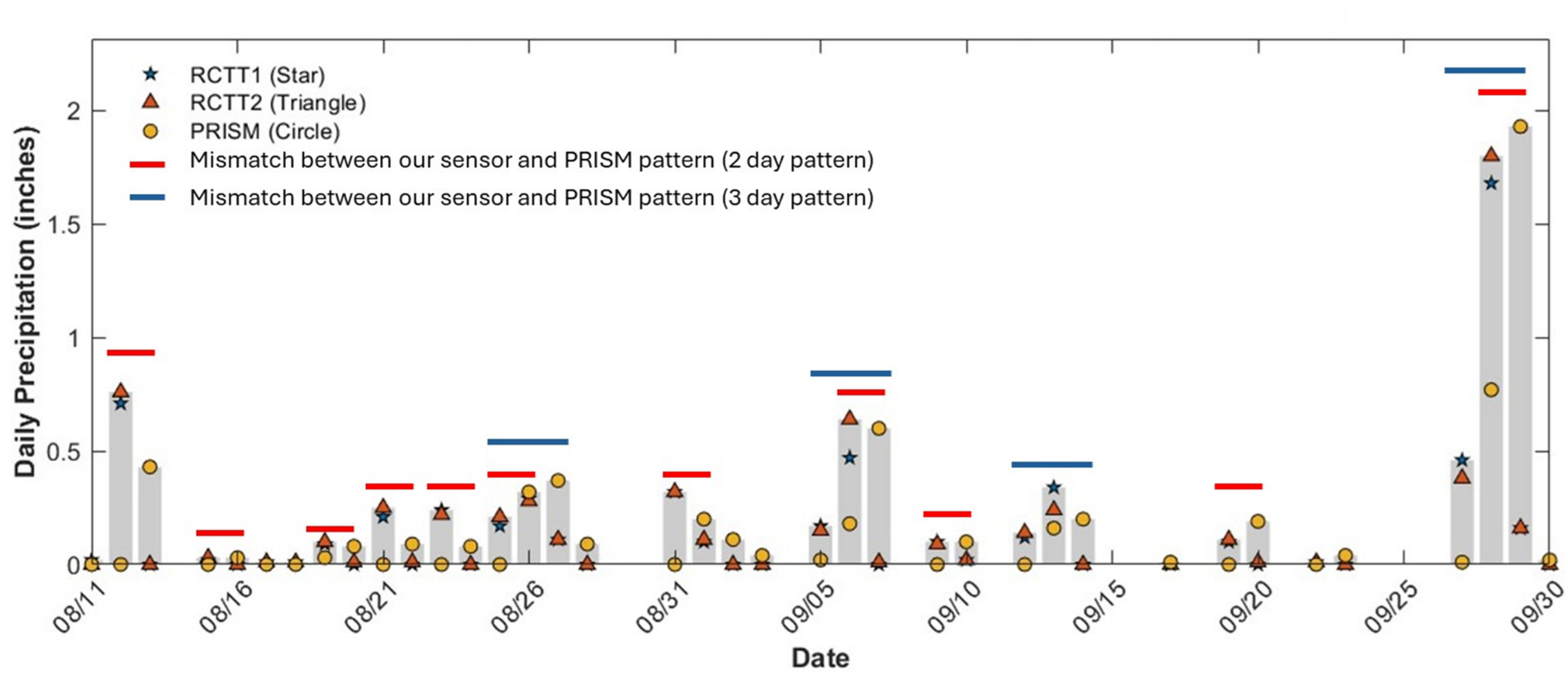

**Figure 11: Daily precipitation comparison between RCTT1, RCTT2, and PRISM for August-September 2025, showing overall agreement with instances of two-day (red) and three-day (blue) temporal mismatches between sensor readings and PRISM estimates.**

The red lines highlight a two-day mismatch where LEWIS 6 sensors detected rainfall earlier than the PRISM dataset. In these cases, rainfall peaks appeared in the sensor data one day before the PRISM model indicated precipitation, suggesting that the model's temporal smoothing may delay short-duration events. Blue lines denote a three-day mismatch pattern, in which PRISM's rainfall peaks lagged by two days compared to LEWIS 6 sensor readings. This occurred when localized rainfall was detected immediately by the on-site sensors but appeared later and spread over multiple days in the gridded PRISM output. The difference can likely be attributed to PRISM's interpolation process, which averages data across surrounding grid cells and elevation zones, reducing temporal precision during convective rainfall episodes.

### *3.4.2. RC01 and RC02 vs PRISM*

Figure 12 compares daily precipitation data recorded by two LEWIS 6 field sensors (RC01 and RC02) with PRISM model estimates from August to September 2025. Gray bars represent the highest daily rainfall among the three data sources, while the red and blue lines highlight periods of temporal mismatches between the sensor observations and PRISM estimates. A total of five red lines indicate two-day mismatches, where the LEWIS 6 sensors detected rainfall one day earlier than the PRISM model. Additionally, three blue lines correspond to three-day mismatches, where PRISM's rainfall peaks occurred approximately two days after the LEWIS 6 sensor events. Overall, while the PRISM model effectively captured broader rainfall trends, LEWIS 6 sensors showed superior temporal sensitivity, allowing precise identification of localized and short-lived rainfall events in near real-time.

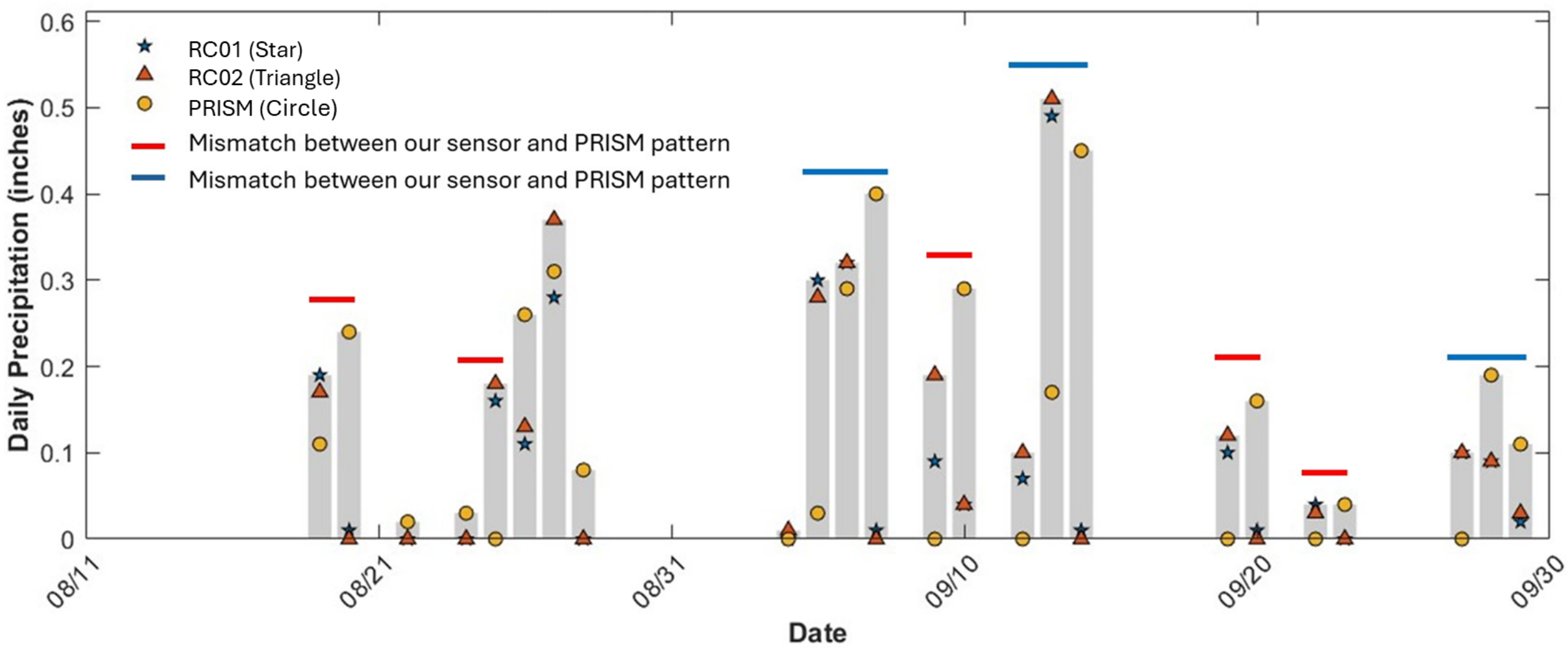

**Figure 12. Daily precipitation comparison between RC01, RC02, and PRISM for August-September 2025, showing general alignment in rainfall trends with instances of two-day (red) and three-day (blue) temporal mismatches between the LEWIS 6 sensor readings and PRISM model estimates.**

Although PRISM provides an essential large-scale rainfall estimation tool and is widely applied in climatology, water resource planning, and hazard assessment, its accuracy is inherently limited by the mathematical interpolation and signal processing techniques used to derive precipitation from sparse ground observations and remotely sensed inputs. These methods often struggle to capture highly localized convective storms, rapid intensity changes, and terrain-driven microclimate effects that are common in regions like New Mexico. As a result, while PRISM performs well for long-term climatological evaluations and regional trend analysis, it may underestimate or over-smooth real precipitation values during short-duration events. These limitations highlight the necessity for a dense network of ground-based rainfall sensors that can provide real-time and fine-scale validation data to improve model accuracy and support better forecasting and hazard preparedness.

### *3.5. Sub-Daily Precipitation Performance*

Figure 13 shows a map of cumulative precipitation based on NEXRAD data for an intense storm event that occurred during the study period (September 28-29, 2025). According to radar-derived precipitation estimates, in excess of 2 inches of rain fell over a 12-hour period just east of the city of Albuquerque in the area of LEWIS 6 stations RCTT1 and RCTT2, while stations RC01 and RC02 in Albuquerque received no rainfall. The chart included at the bottom of Figure 13 displays the cumulative precipitation time series measured by sensor RCTT1 (solid line), compared to the time series for the same location extracted from NEXRAD data (dotted line). Remotely-sensed precipitation data can provide valuable information about the spatial heterogeneity of rainfall events; however, the comparison of time series illustrates a challenge with this data source: radar-derived rainfall estimates are often biased (Baig et al., 2025; Ouatiki et al., 2023; Yaswanth et al., 2023). For the September 28/29 storm event, the difference between LEWIS 6 and NEXRAD was 0.48 inches (or 29%) over the 12-hour storm duration. Rain sensors such as the LEWIS 6 system can provide crucial information for systematic bias correction and substantially improve precipitation estimates based on remotely sensed data.

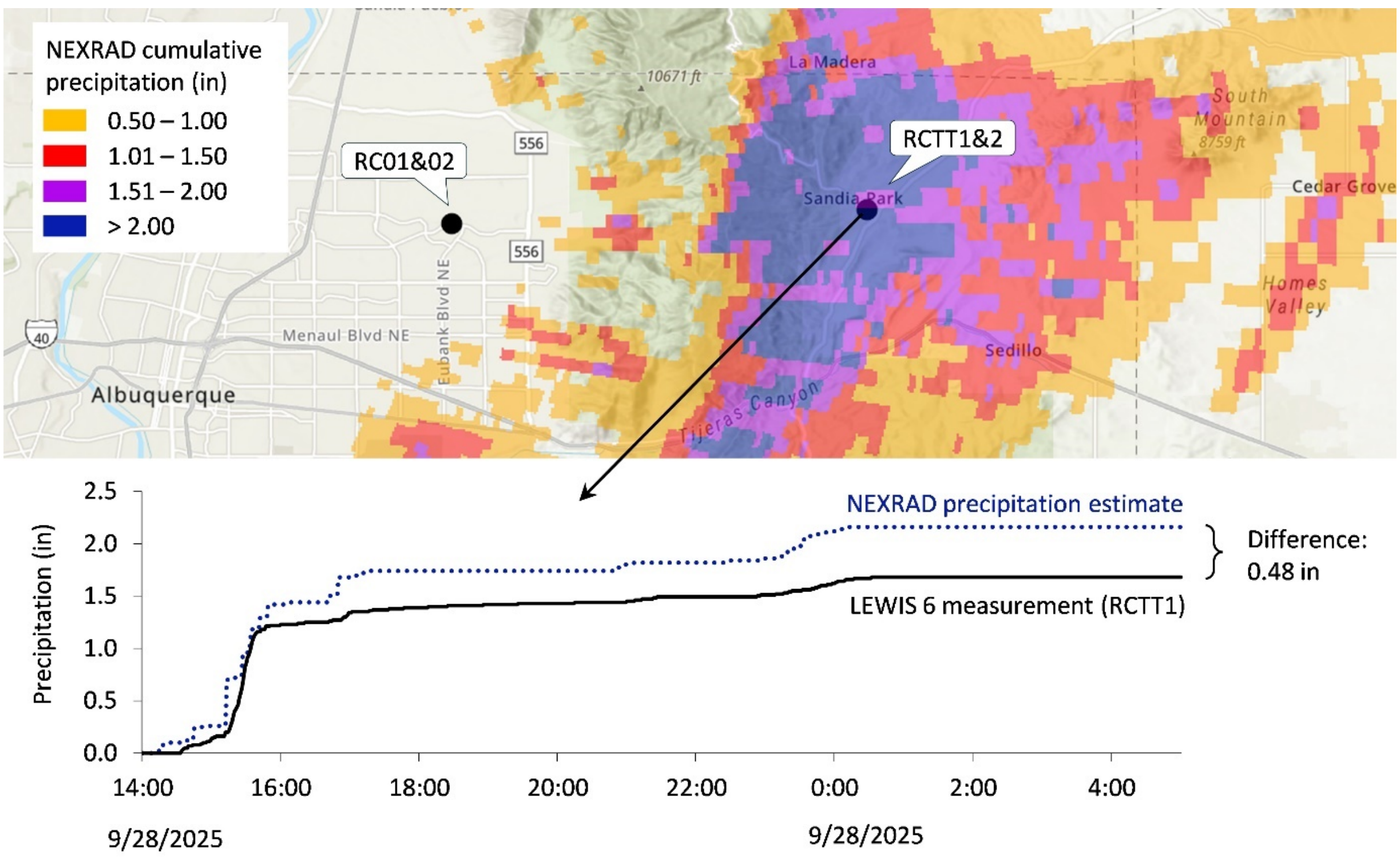

**Figure 13. Map showing NEXRAD-based cumulative precipitation estimates (top) and precipitation time series (bottom) derived from NEXRAD data (dotted line) compared to measurements by a LEWIS 6 sensor (solid line) for a summer thunderstorm that occurred on September 28-29, 2025.**

Conventional rainfall monitoring systems and regional precipitation products provide essential hydrological observations but face limitations in capturing localized rainfall variability at fine spatial and temporal scales. Ground-based instruments often require continuous power supply, periodic calibration, and regular maintenance, which restricts deployment density in remote or resource-limited environments (Posudin, 2014). Similarly, radar-and satellite-derived datasets provide broad spatial coverage but rely on interpolation and reflectivity-based estimation methods that may introduce systematic bias and smooth short-duration events (Joe, 1996; Nirala and Cracknell, 2002). High installation and operational costs further contribute to sparse monitoring networks and observational gaps. Consequently, localized storm dynamics may remain underrepresented in conventional datasets. This study demonstrates that the LEWIS 6 system offers a low-cost, energy-autonomous solution for accurate point precipitation measurement, enabling dense ground-based monitoring, particularly in remote locations and budget-constrained applications.

## 4. Conclusion

This study introduced the LEWIS 6 rainfall monitoring system, a low-cost, solar-powered IoT platform designed for autonomous, real-time precipitation measurement and wireless data transmission. Validation against both the PRISM gridded precipitation data and the CoCoRaHS community-based gauge network revealed strong agreement in overall rainfall trends across multiple temporal scales. In particular, the co-located LEWIS 6 sensors consistently exhibited the lowest variability and RMSE values during daily, weekly, and monthly comparisons, indicating stable sensor performance and high measurement repeatability under field conditions.

Although PRISM effectively captured regional precipitation patterns and broader seasonal accumulation trends, its relatively coarse spatial resolution introduced smoothing effects and temporal averaging that reduced its ability to accurately capture short-duration or highly localized storm events. CoCoRaHS observations provided important point-scale ground-truth reference for validation; however, some variability among gauge measurements was expected due to differences in local rainfall distribution, site exposure, and observational uncertainty. In contrast, the LEWIS 6 system successfully detected localized rainfall peaks and rapid storm event dynamics, highlighting the value of dense in-situ sensing networks for improving temporal resolution and supporting calibration of regional precipitation products.

Comparison of LEWIS 6 measurements with NEXRAD-derived precipitation estimates during an intense summer thunderstorm further illustrated the important observational gap that systems such as the LEWIS 6 can address by identifying and reducing systematic bias in remotely sensed precipitation products. These findings suggest that integrating distributed IoT-based precipitation sensors with regional datasets such as PRISM, NEXRAD, and community observation networks such as CoCoRaHS can provide a more comprehensive and resilient rainfall monitoring framework.

Overall, the integration of high-resolution IoT sensing, regional gridded datasets, radar-based precipitation products, and community-scale observations offers substantial potential for improving precipitation monitoring and hydrologic decision-making. Such integrated monitoring systems can enhance early flood detection, drought assessment, watershed management, and climate-resilient hydrologic modeling by providing both localized rainfall precision and broader regional context. Future work will focus on expanding LEWIS 6 network deployment, incorporating advanced analytics and machine-learning approaches, and developing cloud-based visualization and alert platforms to support real-time environmental monitoring and decision support applications.

## Author Contributions:

**Mohammad Solaiman:** Data curation, Formal analysis, Investigation, Methodology, Software, Validation, Visualization, Writing – original draft, Writing – review & editing; **Ronan Reza:** Visualization, Writing – review & editing; **Su Zhang:** Conceptualization, Funding acquisition, Methodology, Formal analysis, Writing – review & editing, Visualization; **Gerhard Schoener:** Writing – review & editing, Visualization; **Fernando Moreu:** Conceptualization, Formal analysis, Funding acquisition, Investigation, Methodology, Project admin, Resources, Software, Supervision, Validation, Visualization, Writing – review & editing

## Declaration of competing interest

The authors declare that they have no known competing financial interests or personal relationships that could have appeared to influence the work reported in this paper.

## Acknowledgement

This work is funded by the Regional Development Corporation of New Mexico, exploring development of new knowledge of new technologies that can benefit Northern New Mexico (2024-2026); the National Science Foundation (NSF), Division of Computer and Network

Systems (CNS), Directory for Computer and Information Science and Engineering (CISE), Smart and Connected Communities, CIVIL Challenge, Grant/Award Number: 2133334; NSF Division of Information & Intelligent Systems (IIS) CISE, Hardening the Data Revolution DSC, Grant/Award Number: 2123346; and the U.S. Department of Transportation (Award No. 69A3552348306) through the USDOT Southern Plains Transportation Center (SPTC). The authors recognize the support from Tinkertown Museum and Carla Ward for granting us permission to using their property as part of this project and their involvement in selecting sites and providing input in the hardware and software of the LEWIS system, data collection, and sensor locations. The support of the following individuals is highly appreciated: Dr. Jiwi Chong, Hayden Lammons, Ronan Reza, Gavin DeBerry, Guillermo Toledo, Isaac May, and Bryan Rodriguez.